\documentclass[rnote,structabstract]{aa} % for the research notes
\usepackage{graphicx}
\usepackage{txfonts}
\usepackage{natbib}
\bibpunct{(}{)}{;}{a}{}{,} 

\begin{document}
    \title{Groups and clusters of galaxies in the SDSS DR8}
    \subtitle{Value-added catalogues}
    \author{E. Tempel\inst{1,}\inst{2}
    \and
    E. Tago\inst{1}
    \and
    L.~J. Liivam\"agi\inst{1,}\inst{3}
    }

    \institute{Tartu Observatory, Observatooriumi~1, 61602 T\~oravere, Estonia\\
        \email{elmo@aai.ee}
    \and
    National Institute of Chemical Physics and Biophysics, R\"avala pst 10, Tallinn 10143, Estonia
    \and
    Institute of Physics, Tartu University, T\"ahe~4, 51010 Tartu, Estonia
    }

   \date{Received December 20, 2011; accepted February 19, 2012}

% \abstract{}{}{}{}{} 
% 5 {} token are mandatory
 
  \abstract
   {}
   {We intend to compile a new galaxy group and cluster sample of the latest available SDSS data, adding several parameter for the purpose of studying the supercluster network, galaxy and group evolution, and their connection to the surrounding environment.}
   {We used a modified friends-of-friends (FoF) method with a variable linking length in the transverse and radial directions to eliminate selection effects and to find reliably as many groups as possible. Using the galaxies as a basis, we calculated the luminosity density field.}
   {We create a new catalogue of groups and clusters for the SDSS data release~8 sample. We find and add environmental parameters to our catalogue, together with other galaxy parameters (e.g., morphology), missing from our previous catalogues. We take into account various selection effects caused by a magnitude limited galaxy sample. Our final sample contains 576493 galaxies and 77858 groups. The group catalogue is available at \texttt{http://www.aai.ee/$\sim$elmo/dr8groups/} and from the Strasbourg Astronomical Data Center (CDS).}
   {}


   \maketitle
%
%________________________________________________________________

\section{Introduction}

    Observations of the local Universe have shown that basically all galaxies are located in groups -- it is their natural environment. Groups and clusters of galaxies form the basic building blocks of the Universe. Therefore, it is essential to extract groups of galaxies from galaxy surveys, and their study can provide new understanding of the evolution of galaxies, of the large-scale structure, and of the underlying cosmological model.
   
    In our previous papers \citep{Tago:08,Tago:10} we have extracted groups from the SDSS DR5 and DR7 samples, respectively. In these papers we have given an extensive review of papers dedicated to group search methods and of the published group catalogues. In this introduction we present only a short review of the studies of galaxy groups.
        
    During the last decade, several group catalogues that use spectroscopic redshifts have been published, either based on the 2dFGRS \citep{Eke:04, Yang:05, Tago:06}, or on earlier releases of the SDSS \citep{Einasto:03, Merchan:05, Zandivarez:06, Berlind:06, Berlind:09, Yang:07,Koester:07}. However, similar algorithms used to compile these catalogues have yielded groups of galaxies with rather different statistical properties. 
    
    Several authors have recently compiled group catalogues up to the redshift 0.6: GAMA (Galaxy And Mass Assembly) by \citet{Robotham:11} is a galaxy group catalogue based on the SDSS target catalogue; \citet{Farrens:11} derived a catalogue, based on the LRGs and QSOs in the 2dF-SDSS surveys. Using photometric redshifts several group/cluster catalogues have been compiled \citep[e.g.][]{Gal:09,Szabo:11}. \citet{Hao:10} applied a Gaussian mixture BCG algorithm to the SDSS DR7 data and assembled a photometric group catalogue up to the redshift 0.55. Using spectroscopic redshifts, \citet{Knobel:09} compiled the deepest group catalogue so far, reaching up to the redshift 1 in the zCOSMOS field. In addition, catalogues of rich galaxy clusters have been created by \citet{Miller:05}, \citet{Aguerri:07}, and \citet{Popesso:07}. We discuss and compare some of these catalogues in a separate paper.

    The papers dedicated to group and cluster search use a wide range of both sample selection methods as well as cluster search methods and parameters. The choice of the methods and parameters depends on the goal of the catalogue. For example, while \citet{Weinmann:06} searched for compact groups in the SDSS DR2 sample, applying strict criteria in the friend-of-friend (FoF) method, then \citet{Berlind:06} applied the FoF method to the volume-limited samples of the SDSS with the goal to measure the group multiplicity function and to constrain dark matter haloes. Hence, the parameters of the algorithm depend on the goal of the study.
   
    Our goal is to generate an up-to-date catalogue of groups and clusters for large-scale structure studies. The catalogue is based on the SDSS data release~8 (DR8). Since the SDSS spectroscopic main sample basically did not change from DR7 to DR8, we use exactly the same group finding algorithm and parameters as described in \citet{Tago:10}, yielding a group sample with similar properties. The photometry of the galaxies in DR8 has been reprocessed, yielding more accurate luminosities. This is important for detailed photometric galaxy modelling (Tempel et al. in prep).
   
    Compared to our previous catalogues, we added several additional descriptors, including environmental parameters (both local and global) and morphology. For the DR7, such data have been already used in several papers: the environment and morphology have been used to study the environmental effects on galaxy evolution \citep{Tempel:11a}; global environments have been used to extract superclusters from the cosmic network \citep{Liivamagi:10}. The present catalogue, based on the DR8, has already been used to compare the local and global environments of galaxies (Lietzen et al. in prep), to study the structure of rich groups \citep{Einasto:12}, and to study the photometric structure of galaxies (Tempel et al. in prep).
   
    The paper is organised as follows. The data used are described in Sect.~\ref{sect:data}. Section~\ref{sect:cat} gives a brief overview of the group finding algorithm used, together with a short comparison with our DR7 catalogue. In Sect.~\ref{sect:dens} we describe our method to calculate the luminosity density field. In Sect.~\ref{sect:param} we describe the additional galaxy and group parameters. All the parameters in the resulting catalogue are described in Appendix~\ref{app:cat}. The catalogue can be downloaded from \texttt{http://www.aai.ee/$\sim$elmo/dr8groups/} or from the Strasbourg Astronomical Data Center (CDS)\footnote{Galaxy and group/cluster tables will be available at the CDS via anonymous ftp to cdsarc.u-strasbg.fr (130.79.128.5) or via http://cdsweb.u-strasbg.fr/cgi-bin/qcat?J/A+A/}.
    
    Throughout this paper we assume the following cosmology: the Hubble constant $H_0 = 100\,h\,\mathrm{km\,s^{-1}Mpc^{-1}}$, the matter density $\Omega_\mathrm{m}=0.27$ and the dark energy density $\Omega_\Lambda=0.73$.

%__________________________________________________________________
\section{SDSS data}\label{sect:data}

\begin{figure}
   \centering
   \includegraphics[width=88mm]{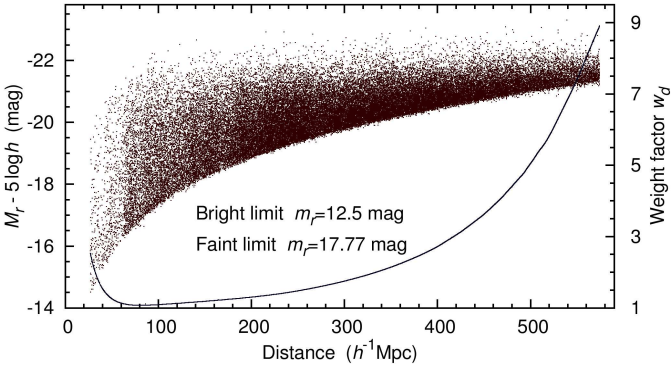}
   \caption{The distance versus the absolute magnitude of the galaxies. The faint magnitude limit is fluctuating due to the $k$-correction. The bright magnitude limit $m_r=12.5$ (used for the weight factor) affects only the nearby region $d<60\,h^{-1}$Mpc. \emph{The solid line} shows the weight factor $W_d$ at a given distance (see Sect.~\ref{sect:dens} for more information).}
   \label{fig:distmag}
\end{figure}

    Our present catalogue is based on the SDSS DR8 \citep{Aihara:11}. We used only the main contiguous area of the survey (the Legacy Survey). The galaxy data were downloaded from the Catalog Archive Server (CAS) of the SDSS. The primary selection was based on the \texttt{specphotoall} table in the CAS and we used only those objects that were classified as galaxies. Since the spectroscopic galaxy sample is complete only up to the Petrosian magnitude $m_r=17.77$ \citep{Strauss:02}, we select that as the lower magnitude limit of our sample. Actually, that limit was applied after the Galactic extinction correction was used, yielding an uniform extinction-corrected sample. Initially, we set no upper magnitude limit to our catalogue. However, since the SDSS sample is incomplete for bright objects due to the saturation of CCDs, we used the limit $m_r=12.5$ for the luminosity function and for the weight factor calculations. The bright limit affects only the nearby regions $d<60\,h^{-1}$Mpc (see Fig.~\ref{fig:distmag}).
    
    However, the sample is still affected by fibre collisions -- the minimum separation between spectroscopic fibres is 55\,\arcsec. For this reason, about 6 per cent of galaxies in the SDSS are without observed spectra. In \citet{Tago:08} we showed that it does not generate any appreciable effects, when using our group-finding algorithm. In the present paper we track the missing galaxies and add a flag to the galaxy/group with a neighbour(s) missing from the redshift catalogue. According to \citet{Patton:08} and \citet{Ellison:08}, 67.5\% of close pairs with angular separations below 55\,\arcsec are missing due to fibre collision. This reduces the number of galaxy pairs in our group catalogue. Using the missing galaxies, we can estimate the number of missing pairs. In the SDSS sample, the absent galaxies are more likely to reside in groups and there are only 4\% of single galaxies which have a missing companion. Furthermore, only 60\% of them have redshift close to the neighbour's \citep{Zehavi:02}. As a result, the estimated amount of missing pairs in our catalogue is about 8\%.
    
    The galaxy sample, downloaded from the CAS, needs further checking, since it includes duplicate entries and in some cases objects that are not galaxies at all (but still classified as galaxies in the CAS). To obtain a clean sample of galaxies, we firstly excluded duplicate entries, using the redshifts and angular distances between galaxies. We also used the SDSS Visual Tool to examine the cases, where duplication was unclear (merging or visually extremely close galaxies). We also examined all of the 1000 brightest galaxies in the remaining sample and excluded the entries, which were not galaxies: in most cases these objects were oversaturated stars or other artefacts. This step was crucial, since for the luminosity density field, the brightest objects that are not galaxies can cause the biggest uncertainties in the final estimates. Additionally, we visually checked the galaxies, which had unphysical colours, and excluded all spurious objects.
    
    After correcting the redshifts relative to the motion in respect of the CMB, we put the lower and upper distance limits to $z=0.009$ and $z=0.2$, respectively. The lower limit was set to exclude the local supercluster, and the upper limit was chosen, since at larger distances the sample becomes very diluted. As a result, after all the limits and exclusions, our final sample includes 576493 galaxies.
    
    The apparent magnitude $m$ was transformed into the absolute magnitude $M$ according to the usual formula
\begin{equation}
    M_\lambda = m_\lambda - 25 -5\log_{10}(d_L)-K,
\end{equation} 
    where $d_L$ is the luminosity distance in units of $h^{-1}$Mpc, $K$ is the $k$+$e$-correction, and the index $\lambda$ refers to the $ugriz$ filters. The $k$-corrections were calculated with the \mbox{KCORRECT\,(v4\_2)} algorithm \citep{Blanton:07} and the evolution corrections were estimated, using the luminosity evolution model of \citet{Blanton:03}: $K_e=c\!\cdot\! z$, where $c=-4.22,\,-2.04,\,-1.62,\,-1.61,\,-0.76$ for the $ugriz$ filters, respectively. The magnitudes correspond to the rest-frame (at the redshift $z=0$).

\begin{figure}
   \centering
   \includegraphics[width=88mm]{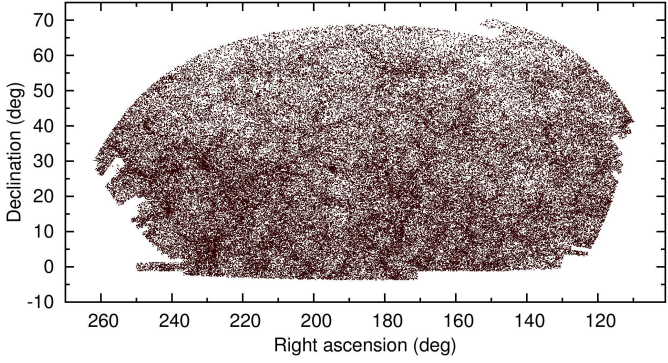}
   \caption{The SDSS contiguous sample area in the equatorial coordinates. The sky coverage is 7221 square degrees.}
   \label{fig:crd}
\end{figure}

    Figure~\ref{fig:crd} shows the sky distribution of galaxies in the equatorial coordinates for our sample, covering 7221 square degrees in the sky \citep{Martinez:09}. Figure~\ref{fig:distmag} shows the distance versus absolute magnitude plot. The flux-limited selection is well seen: further away, only the brightest galaxies are observed.

%__________________________________________________________________
\section{Construction of the group catalogue}\label{sect:cat}

    The details of our group finding algorithm are explained in  detail in \citet{Tago:08,Tago:10}. In this research note we give only a brief outline of the method used.
    
    One of the most conventional methods to search for groups of galaxies is cluster analysis that was introduced in cosmology by \citet{Turner:76}. This method was named friends-of-friends (FoF) by \citet{Press:82}. With the FoF method, galaxies are linked into systems, using a certain linking length (or neighbourhood radius). Choosing the right linking length is rather complicated. In most cases, the linking length is not constant, but varies with distance and/or other parameters.
    
    Our experience shows that the choice of the linking length depends on the goals of the specific study. In our group catalogue, our goal is to obtain groups to estimate the luminosity density field and to study the properties of the galaxy network. Hence, our goal is to find as many groups as possible, whereas the group properties must not change with distance. In our group definition, we tried to avoid the inclusion of large sections of surrounding filaments or parts of superclusters.

    To find the proper scaling for the linking length with distance, we created a test group catalogue, using a constant linking length. Then we selected in the nearby volume ($d < 200\,h^{-1}\mathrm{Mpc}$) all groups with more than 20 members. Assuming that the group members are all at the mean distance of the group, we determined their absolute magnitudes and peculiar radial velocities. Then we shifted these nearby groups, calculating the parameters of the groups (new $k$+$e$-corrections and apparent magnitudes), as if the groups were located at larger distances. As with the increasing distance more and more fainter members of groups fall outside the observational window of apparent magnitudes, the group membership changes. We then determined new properties of the groups -- their multiplicities, characteristic sizes, rms velocities, and number densities. We also calculated the minimum FoF linking length necessary to keep the group together at this distance. Determining the mean values of the group linking lengths, we found that the linking length in our group finding algorithm increases moderately with distance. A good approximation of the scaling law for the linking length ($d_{LL}$) is the arctan function
    \begin{equation}
        d_{LL}(z)=d_{LL,0}\left[1+a\arctan(z/z_\star)\right],
    \end{equation}
    where $d_{LL,0}$ is the value of linking length at the initial redshift; $a$ and $z_\star$ are the parameters. For the DR7 groups we find the parameter values $a = 1.00$ and $z_\star = 0.050$. The ratio of the radial to the transversal linking lengths was 10 (if the radial linking length in km\,s$^{-1}$ is transformed into a formal ``distance'' in $h^{-1}$Mpc). We used the following initial linking length values: 250\,km\,s$^{-1}$ for the radial length and 0.25\,$h^{-1}$Mpc for the transversal length. We use the same values for the DR8 data. Higher initial values would lead to including galaxies from neighbouring groups and filaments; lower values exclude the fastest members in intermediate richness groups. The selected parameters lead to reasonable group properties.

    Our final group catalogues are rather homogeneous. The group richnesses, mean sizes and velocity dispersions practically do not depend on their distance. The homogeneity of our catalogues have been tested also by other authors. For example, \citet{Tovmassian:09} select poor groups from our SDSS catalogues and conclude that the main parameters of our groups are distance independent and well suited for statistical analysis.
    
    As a final result, the group catalogue includes 77858 groups with two or more members. Table~\ref{table:gr_rich} shows the numbers (and fractions) of groups and galaxies in different group richness bins. Almost half of the galaxies in our sample (46\%) belong to a group and 10\% of galaxies belong to groups with ten or more members. Most of our groups (60\%) are groups with two (observable) members and 21\% of groups have four or more members.
    
\begin{table}
\caption{The numbers of groups ($N_\mathrm{gr}$) and galaxies ($N_\mathrm{gal}$) in different group richness ($N_\mathrm{rich}$) bins.}
\label{table:gr_rich}
\centering
\begin{tabular}{l r l r l}
\hline\hline
Group richness & $N_\mathrm{gr}$ & Fraction\tablefootmark{a} & $N_\mathrm{gal}$ & Fraction\tablefootmark{b} \\
\hline
All     & 77858 & 100.0 & 265578 & 46.1 \\
$N_\mathrm{rich}\geq 3$ & 30515 & \phantom{0}39.2 & 170892 & 29.6 \\
$N_\mathrm{rich}\geq 4$ & 16358 & \phantom{0}21.0 & 128421 & 22.3 \\
$N_\mathrm{rich}\geq 5$ & 10150 & \phantom{0}13.0 & 103589 & 18.0 \\
$N_\mathrm{rich}\geq 6$ & 7078  &  \phantom{00}9.1 & 88229  & 15.3 \\
$N_\mathrm{rich}\geq 8$ & 4078  &  \phantom{00}5.2 & 69060  & 12.0 \\
$N_\mathrm{rich}\geq 10$ & 2686 &  \phantom{00}3.4 & 57369  & 10.0 \\
$N_\mathrm{rich}\geq 15$ & 1278 &  \phantom{00}1.6 & 41091  & \phantom{0}7.1 \\
$N_\mathrm{rich}\geq 20$ & 772  &  \phantom{00}1.0 & 32646  & \phantom{0}5.6 \\
$N_\mathrm{rich}\geq 30$ & 412  &  \phantom{00}0.5 & 24169  & \phantom{0}4.2 \\
$N_\mathrm{rich}\geq 40$ & 233  &  \phantom{00}0.3 & 18145  & \phantom{0}3.1 \\
$N_\mathrm{rich}\geq 50$ & 147  &  \phantom{00}0.2 & 14349  & \phantom{0}2.5 \\
$N_\mathrm{rich}\geq 75$ & 66   & \phantom{00}0.08 & 9528   & \phantom{0}1.6 \\
$N_\mathrm{rich}\geq 100$ & 35  & \phantom{00}0.04 & 6965   & \phantom{0}1.2 \\
\hline
$N_\mathrm{rich}= 2$ & 47343  & \phantom{0}60.8 & 94686   & 16.4 \\
$3\leq N_\mathrm{rich}\leq 4$ & 20365  & \phantom{0}26.2 & 67303   & 11.7 \\
$5\leq N_\mathrm{rich}\leq 9$ & 7464   &  \phantom{00}9.6 & 46220   & \phantom{0}8.0 \\
$10\leq N_\mathrm{rich}\leq 29$ & 2274   &  \phantom{00}2.9 & 33200   & \phantom{0}5.6 \\
\hline
\end{tabular}
\tablefoot{
\tablefoottext{a}{Fraction of groups in per cent.}
\tablefoottext{b}{Fraction of galaxies in groups (per cent).}
}
\end{table}

%__________________________________________________________________
\subsection{Comparison with the DR7 group catalogue}

\begin{figure}
   \centering
   \includegraphics[width=80mm]{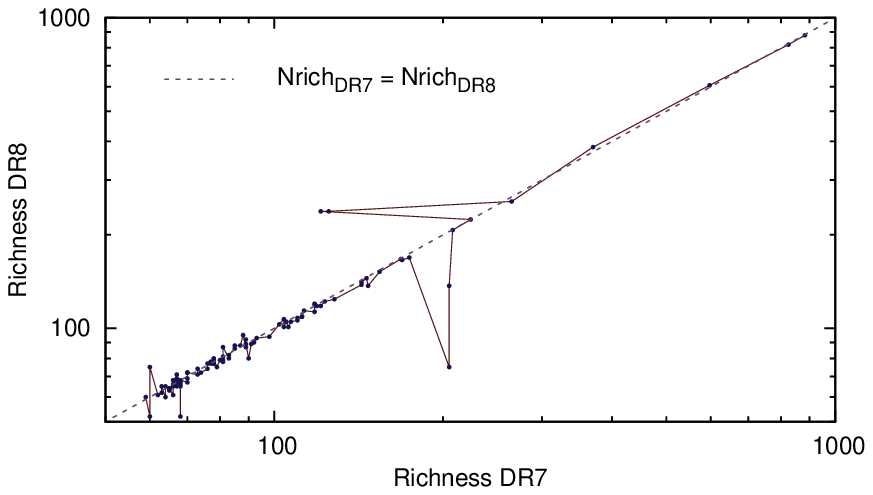}
   \caption{Comparison of the richness of the 101 richest groups in the DR7 and DR8. There are no systematic deviations from
    the line $\mathrm{Nrich}_\mathrm{dr7} = \mathrm{Nrich}_\mathrm{dr8}$.}
   \label{fig:rich7v8}
\end{figure}

    Considering  that the SDSS DR7 and DR8 galaxy catalogues are different (due to a reprocessing of the photometry of all earlier releases and due to different criteria for sample cleaning) it is useful to compare the respective group catalogues and to determine how different are the groups identified as ``the same'' in both catalogues. We have compared the richness of the 101 richest groups in both catalogues. As shown in Fig.~\ref{fig:rich7v8}, there are no systematic trends. A few individual fluctuations can only be seen: in one case, a DR7 group is split into two groups in the DR8 and in one case the two DR7 groups are merged into one group in the DR8.

    Compared to our previous DR7 catalogue, there are about 1000 groups less in the present catalogue. The biggest difference is in the number galaxy pairs, the present catalogue contains 600 pairs less than the DR7 catalogue. The number of groups with 30 and more members is practically the same in both catalogues. We also matched the groups in the DR7 and present catalogues, using the group richness and the brightest galaxies in groups. We were able to match 94\% of the groups, where the groups in DR7 and DR8 have at least 90\% common galaxies. The remaining 6\% are mostly smaller groups and/or are split into two groups or merged into one group.

\begin{figure}
   \centering
   \includegraphics[width=88mm]{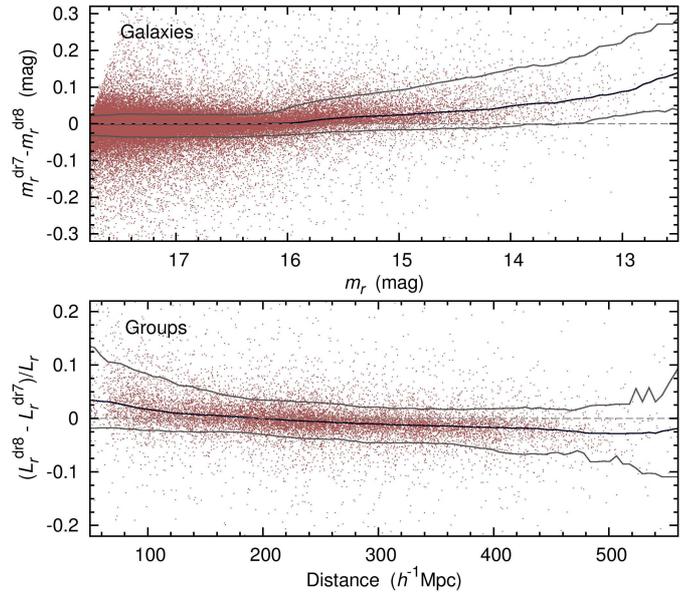}
   \caption{\emph{The upper panel} shows the differences between the SDSS DR7 and DR8 Petrosian galaxy magnitudes as a function of galaxy luminosity. \emph{The lower panel} shows the relative difference between the SDSS DR7 and DR8 group luminosities as a function of distance for groups with four and more members, which were identified as the same in the DR7 and DR8 samples. Solid lines show the 0.1, 0.5, and 0.9 quantiles, respectively.}
   \label{fig:comp_lum}
\end{figure}

    Since the main change between the DR7 and DR8 data concerns the photometry, we evaluated how the group luminosities differ when compared to our previous catalogue. The upper panel in Fig.~\ref{fig:comp_lum} shows the magnitude differences between the DR7 and DR8 for galaxies. The reprocessed photometry increases the brightness of luminous galaxies, mainly due to a better estimate of the sky background. The bottom panel in Fig.~\ref{fig:comp_lum} shows the relative difference in group luminosities for the DR7 and DR8 data as a function of distance. In this plot, only the groups with four and more members and which were identified as the same in the DR7 and DR8 samples are shown. We see a slight trend with distance: nearby groups in the DR8 are a few per cent more luminous and more distant groups less luminous than the groups in the DR7. The same trend is also visible for galaxies and is caused by the reprocessed photometry. Figure~\ref{fig:comp_lum} shows that for majority of the groups, the difference in the group luminosity between the DR7 and DR8 catalogues is less than 5 per cent.

    We also compared other properties of groups in the DR7 and DR8 and our analysis confirms that the properties of the groups in both catalogues are very similar. In order not to overcrowd the paper with figures, we do not show these comparisons here. The basic properties of the groups in the DR7 are described in \citet{Tago:10}, for the DR8, the properties are similar. 

%======================================================================
\section{Estimating the environmental densities}
\label{sect:dens}

\begin{figure}
   \centering
   \includegraphics[width=88mm]{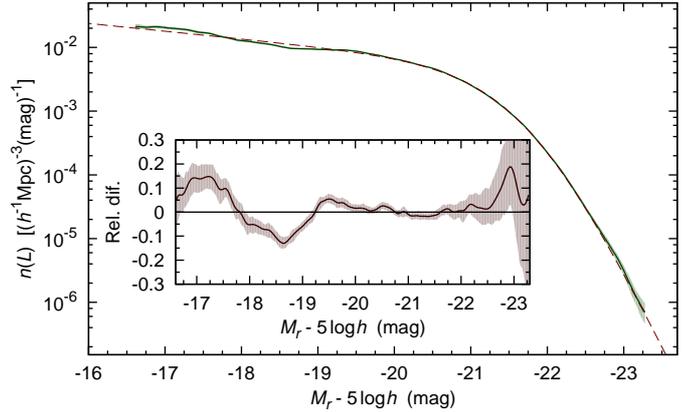}
   \caption{The differential luminosity function in the $r$-band. The dashed line shows the double-power-law fit. The grey area shows the 95\% confidence limits. \emph{The inset panel} shows the relative difference ($n(L)_\mathrm{obs}/n(L)_\mathrm{mod}-1$) between the analytical fit and the numerical estimate.}
   \label{fig:lf}
\end{figure}

    In this group catalogue, we add environmental densities to the galaxies. These densities are important when analysing the influence of local and/or global environments on galaxy evolution.
    
    We calculate the densities as described by \citet{Liivamagi:10}. To calculate the luminosity density field, we need to know the expected total luminosities of groups and isolated galaxies. The primary factor that determines the calculation of group luminosities is the selection effect, present in a flux-limited survey: further away, only the brightest galaxies are seen (see Fig.~\ref{fig:distmag}). To take this into account, we calculated for each galaxy a distance-dependent weight factor $W_d$
    \begin{equation}
        W_d = \frac{\int_0^\infty L n(L)\mathrm{d}L}{\int_{L_1}^{L_2} L n(L)\mathrm{d}L},
    \end{equation}
    where $L_{1,2}=L_\odot 10^{0.4(M_\odot - M_{1,2})}$ are the luminosity limits of the observational window at the distance $d$, corresponding to the absolute magnitude limits of the window $M_1$ and $M_2$; we took $M_\odot=4.64$\,mag in the $r$-band \citep{Blanton:07}. To calculate the magnitudes $M_1$ and $M_2$ we use the average $k\,+\, e$-corrections at a given distance. Due to peculiar velocities, the distances of galaxies are somewhat uncertain; if the galaxy belongs to a group, we use the group distance to determine the weight factor. In the latter equation, $n(L)$ is taken to be the luminosity function in the $r$-band for all galaxies. We used the numerical luminosity function in the regions where the luminosity function was accurately determined, and used the analytical double-power-law approximations only at the bright and the faint end. The distance-dependent weight factor is shown in Fig.~\ref{fig:distmag} as a solid line. In nearby regions, the weight factor increases due to the bright limit of the survey (12.5\,mag). Further away than 400\,$h^{-1}$Mpc, the weight factor increases rapidly due to the increasing number of galaxies with apparent luminosities lower than the luminosity limit of the survey (17.77\,mag).

    To determine the luminosity function, we used the modified $V_\mathrm{max}^{-1}$ weighting procedure with kernel smoothing and varying kernel widths. The luminosity function $n(L)$ (the number density of galaxies) is represented by a sum of kernels centred at the data points:
\begin{equation}
    n(L)=\sum\limits_i\frac{1}{V_\mathrm{max}(L_i)}\frac{1}{a_i}K\left(\frac{L-L_i}{a_i}\right).
\end{equation}
    We use the $B_3(\cdot)$ spline function (Eq.~\ref{eq:b3}) with a width $a$ as the kernel $K(\cdot)$. The kernels are distributions with $K(x)>0,\int K(x)\mathrm{d}x=1$, of zero mean. In the latter equation, the kernel widths depend on the data, $a_i=a(L_i)$; in regions with less data points, we use wider kernels. $V_\mathrm{max}(L)$ is the maximum volume where a galaxy of a luminosity $L$ can be observed in the present survey, and the sum is over all galaxies. This procedure is non-parametric, and gives both the form and true normalisation of the luminosity function. We refer to the \citet{Tempel:11a} for a more detailed description of the procedure.
    
     The resulting luminosity function is shown in Fig.~\ref{fig:lf}, together with the double-power-law fit. We used the double-power-law in the form
\begin{equation}
        n (L) \mathrm{d}L \propto (L/L^{*})^\alpha \left[1 + (L/L^{*})^\gamma\right]^\frac{\delta-\alpha}{\gamma} \mathrm{d}(L/L^{*}), 
        \label{eq:abell}
\end{equation}
    where  $\alpha$ is the exponent at low luminosities $(L/L^{*}) \ll 1$, $\delta$ is the exponent at high luminosities    $(L/L^{*}) \gg 1$, $\gamma$ is a parameter that determines the speed of transition between the two power laws, and $L^{*}$ is the characteristic luminosity of the transition. We find the best parameters to be: $\alpha=-1.305\pm0.009$, $\delta=-7.13\pm0.22$, $\gamma=1.81\pm0.05$, and $M^{*}=-21.75\pm0.05$ (corresponds to $L^{*}$).
    
    To calculate the expected total luminosities of groups, we regard every galaxy as a visible member of a group. For isolated/single galaxies we made an assumption that only the brightest galaxy of the group is visible and therefore the isolated galaxy is also part of some group \citep{Tempel:09}. This assumption is supported by observations of nearby galaxies, which indicate that practically all galaxies are located in systems of galaxies of various size and richness.
    
    Assuming that every galaxy also represents a related group of galaxies, which may lie outside the observational window of the survey, the estimated total luminosity per one visible galaxy is
    \begin{equation}
        L_\mathrm{tot}=L_\mathrm{obs}\cdot W_d,
    \end{equation}
    where $L_\mathrm{obs}$ is the observed luminosity of the galaxy. The luminosity $L_\mathrm{tot}$ takes into account the luminosities of unobserved galaxies and therefore it can be used to calculate the full luminosity density field.

    To determine the luminosity density field, we use a kernel sum:
\begin{equation}
    \ell_{\mathbf{i}} = \frac1{a^3}\sum_\mathrm{gal}  K^{(3)}\left(\frac{\mathbf{r}_\mathrm{gal} -
    \mathbf{r}_\mathbf{i}}{a}\right) L_\mathrm{tot},
    \label{eq:dfield}
\end{equation}
    where $L_\mathrm{tot}$ is the weighted galaxy luminosity, and $a$ -- the kernel scale. For kernel $K(\cdot)$ we use the $B_3$ spline function:
\begin{equation}
    \label{eq:b3}
    B_3(x) = \frac{|x-2|^3 - 4|x-1|^3 + 6|x|^3 - 4|x+1|^3 + |x+2|^3}{12}.
\end{equation}

    The luminosity density field is calculated on a regular cartesian grid generated by using the SDSS $\eta$ and $\lambda$ angular coordinates. This allows an efficient placement of the field in a cube and also a relatively straightforward definition of the sample mask. The field mask is designed to follow the edges of the galaxy sample in the plane of the sky. The contours of the mask are given on Fig.~\ref{fig:maskfc}, and we use constant maximum and minimum distance limits of 55 and 565 $h^{-1}$Mpc.

    The estimates of the luminosity density field near the edges of the survey are biased since we miss the luminosities of the galaxies outside the survey area. For our $B_3(\cdot)$ kernel, the estimate is not affected, if the distance to the edge is twice as large as the smoothing scale. The estimates are fairly reliable also for galaxies with the edge distance larger than one smoothing scale. To control this effect, we added to the catalogue the distance from the edge of the survey for every galaxy. These can be used to select galaxies with unbiased estimates of the environmental luminosity density. The distance can also be used to determine whether a group or cluster is complete or if some of its galaxies could lie outside the survey.

\begin{figure}
   \centering
   \includegraphics[width=85mm]{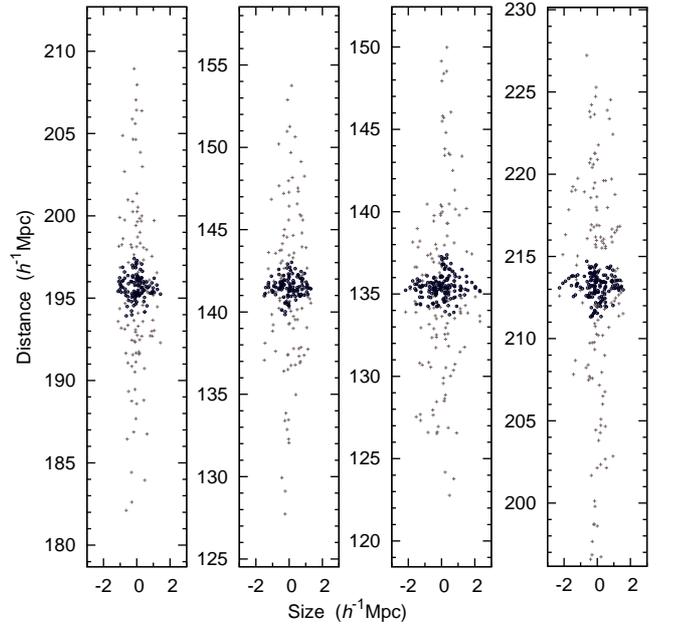}
   \caption{Suppression of finger-of-god redshift distortions for four groups. The $y$-axis shows the distance of a group galaxy in redshift space (\emph{grey points}) and its corrected distance (\emph{black points}). The $x$-axis shows the distance of the galaxy from the group centre in the sky in units of $h^{-1}$Mpc.}
   \label{fig:dr8_sfaris}
\end{figure}

    While calculating the density field, we also suppress the finger-of-god redshift distortions using the rms sizes of galaxy groups in the sky $\sigma_r$ and their rms radial velocities $\sigma_v$ (both in physical coordinates at the location of the group).  For that, we calculate the new radial distances for galaxies $d_{\mbox{gal}}$ as
\begin{equation}
    d_{\mbox{gal}}=d_{\mbox{group}}+\left(d^{\star}_{\mbox{gal}}-d_{\mbox{group}}\right)\frac{\sigma_r}{\sigma_v/H_0},
    \label{eq:distcor}
\end{equation}
    where $d^{\star}_{\mbox{gal}}$ is the initial distance to the galaxy, and $d_{\mbox{group}}$ is the distance to the group centre. For double galaxies, where the extent of the system in the plane of the sky does not have to show its real size (because of projection effects), we demand that its (co-moving) size along the line-of-sight does not exceed the co-moving linking length $d_{LL}(z)$ used to define the system:
\begin{eqnarray}
    d_{\mbox{gal}}&=&d_{\mbox{group}}+\left(d^{\star}_{\mbox{gal}}-d_{\mbox{group}}\right)\frac{d_{LL}(z)}{|v_1-v_2|/H_0},\\
    &&\mbox{if}\;|v_1-v_2|/H_0 > d_{LL}(z).\nonumber
\end{eqnarray}
    Here $z$ is the mean redshift of the double system. If the velocity difference is smaller than that quoted above, we do not change galaxy distances.
    
    The  velocity dispersion $\sigma_v^2$ for groups were calculated with the standard formula
\begin{equation}
    \sigma_v^2 = \frac{1}{(1+z_\mathrm{m})^2(n-1)}\sum\limits^{n}_{i=1}(v_i-v_\mathrm{mean})^2,
\end{equation}
where $v_\mathrm{mean}$ and $z_\mathrm{m}$ are the mean group velocity and redshift, respectively, $v_i$ is the velocity of an individual group member, and $n$ is the number of galaxies with observed velocities in a group.

    In Eq.~(\ref{eq:distcor}), $\sigma_r$ defines the extent of the group in the sky, which is defined as
\begin{equation}
    \sigma_r^2 = \frac{1}{2n(1+z_\mathrm{m})^2}\sum\limits^{n}_{i=1}(r_i)^2,
\end{equation}
where $r_i$ is the projected distance in the sky from group centre (in co-moving coordinates, in units of $h^{-1}$Mpc), and $z_\mathrm{m}$ is the mean group redshift.
    
    We note that such a compression will lead to a better estimate of the density field, but it is unsuitable for a detailed study of individual groups and clusters. Figure~\ref{fig:dr8_sfaris} gives an example of suppressing the finger-of-god redshift distortions for four relatively rich groups. As we see, this procedure makes the galaxy distribution in groups approximately spherical, as intended. We normalise the density field with respect to the mean luminosity density. The mean density is calculated as an average over all density field vertices $\ell_{\mathbf{i}}$ inside the mask. We find the environmental density for all galaxies and groups by linearly interpolating the density field values in neighbouring vertices for the location of the galaxy or the group. The details of the calculation of the luminosity density field can be found in \citet{Liivamagi:10}.

%======================================================================
\section{Galaxy and group parameters added to the catalogue}\label{sect:param}

\subsection{Environmental densities}

    To estimate the environmental densities, we used the method described in Sect.~\ref{sect:dens}. The densities are determined, using the SDSS $r$-band luminosities. Since different smoothing scales represent different environments, we calculated the density field with various smoothing lengths: 1, 2, 4, 8, and 16\,$h^{-1}$Mpc using a 1\,$h^{-1}$Mpc grid. While the smaller smoothing lengths represent the group scales, the larger smoothing lengths correspond to the large-scale environments, to the supercluster-void network.

    For every galaxy in our sample, we find the density field value in the location of the galaxy for all five smoothing scales. For groups, we find the mean density for all the galaxies in the group. These density field values are included in our galaxy/group catalogues. 

\begin{figure}
   \centering
   \includegraphics[width=88mm]{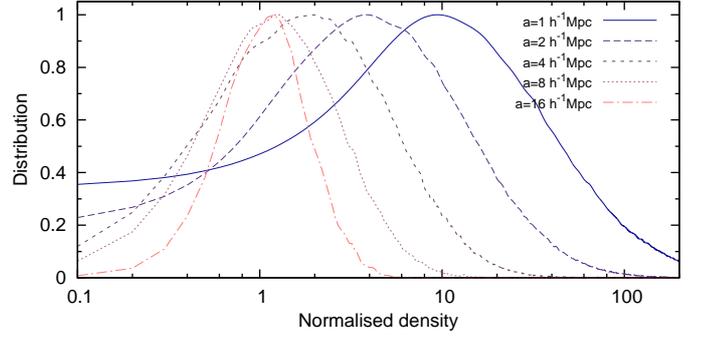}
   \caption{The distribution of the normalised densities for galaxies for various smoothing scales: 1, 2, 4, 8, and 16\,$h^{-1}$Mpc. Note that the maximum shifts toward lower densities up to the scale $a=8\,h^{-1}$Mpc.}
   \label{fig:galdens}
\end{figure}

\begin{figure}
   \centering
   \includegraphics[width=88mm]{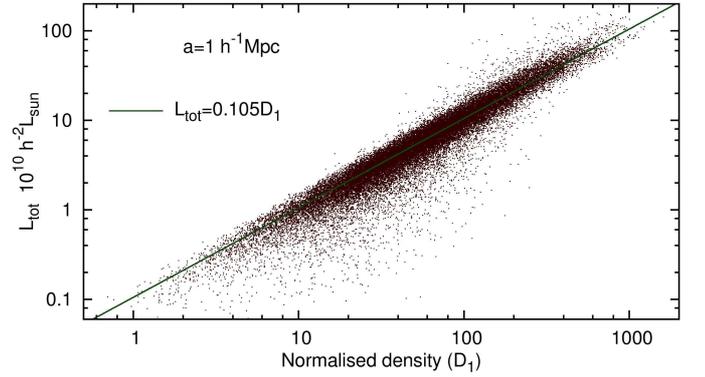}
   \caption{The total luminosity ($L_\mathrm{tot}$) of groups as a function of the normalised density ($D_1$), for the scale $a=1\,h^{-1}$Mpc. The observed group luminosities are multiplied by the weight factor to get the total luminosities. \emph{The solid line} shows a simple linear correlation between these two quantities: $L_\mathrm{tot}=0.105\cdot D_1$.}
   \label{fig:cldens}
\end{figure}

Figure~\ref{fig:galdens} shows the distribution of normalised densities for galaxies for various smoothing scales: 1, 2, 4, 8, and 16\,$h^{-1}$Mpc. The mean density for all smoothing scales is approximately $0.0165\!\times\!10^{10}\,h{L}_{\sun}\mathrm{Mpc}^{-3}$. It is well seen that the maximum shifts towards the mean value when moving toward higher smoothing scales. However, there is no shift after the smoothing scale reaches $a=8\,h^{-1}$Mpc, indicating that a wider smoothing does not reveal any new structures in the density field. The smoothing scale $a=8\,h^{-1}$Mpc was also used to find the largest structures in the cosmic web -- superclusters by \citet{Liivamagi:10}.

Figure~\ref{fig:cldens} shows the normalised density for the smoothing scale $a=1\,h^{-1}$Mpc versus the expected total luminosity of groups. The expected total luminosity is the product of the observed luminosity with the weight factor. There is a clear correlation between this density and the luminosity of groups.

The environmental densities refer to different structures in the cosmic web. The small smoothing scale ($a=1\,h^{-1}$Mpc) describes the group environment and can be therefore used as a local density estimator. The larger smoothing scale ($a=8\,h^{-1}$Mpc) corresponds to the large-scale environment, to the supercluster-void network. Hence, for studying the environmental effects for small and large scales, different density estimators should be used.

In our density field estimation, the mean density is distance independent, since we use a proper weight factor. However, at larger distances we see only the brightest galaxies and the missing luminosity is added only to these locations. Hence, for these distances, the high peaks in the density field are more prominent. The effect is stronger for smaller smoothing scales. Our experience has shown that for distances smaller than 500\,$h^{-1}$Mpc, the densities can be used safely for smoothing scales $a=4\,h^{-1}$Mpc and larger. For smaller smoothing scales, the environmental densities are reliable in the regions, where the weight factor is less than two or three.

\subsection{Galaxy morphology}
\label{sect:morf}

    Galaxy morphology is an important aspect in galaxy evolution studies. For the SDSS sample, the galaxy morphologies have been estimated by the Galaxy Zoo project, yielding a reliable visual classification for the majority of galaxies in SDSS sample. However, the visual classification is subjective and needs to be tested with other methods.
    
    In \citet{Tempel:11a} we carried out a morphological classification of the SDSS galaxies, using various galaxy parameters. This classification takes into account the SDSS model fits, apparent ellipticities (and apparent sizes), and different galaxy colours.
    
\begin{figure}
    \centering
    \includegraphics[width=88mm]{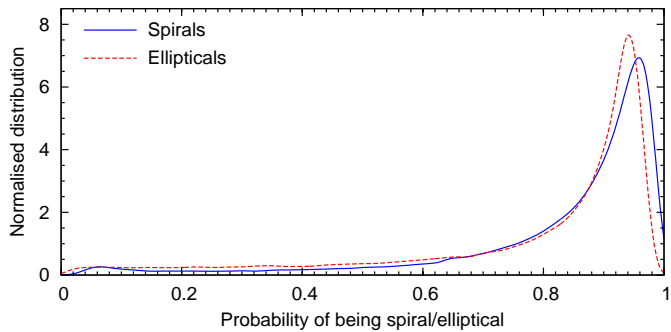}
    \caption{The distribution of the \citet{Huertas-Company:11} marks of early- or late-type galaxies for our classified spirals (blue solid line) and ellipticals (red dashed line).}
    \label{fig:huertas}
\end{figure}

    Recently, \citet{Huertas-Company:11} published a morphological classification of galaxies of the SDSS, based on the Galaxy Zoo data \citep{Lintott:08}. They associate with each galaxy a probability (mark) of being in the four morphological classes: two early-type classes (E and S0) and two late-type classes (Sab and Scd). To compare it with our classification we assign the \citet{Huertas-Company:11} probability of being early- or late-type to our galaxies. In Fig.~\ref{fig:huertas} the distributions of these marks are shown. It is well seen that our classification agrees well with the \citet{Huertas-Company:11} classification.

    In our catalogue we give a flag for the galaxy being a spiral or an elliptical. The flag is given only for these galaxies, where the \citet{Huertas-Company:11} mark is greater than 0.5. Hence, our classification is rather conservative. In our classification about half of the galaxies (45\%) are spirals, about one quarter (26\%) are ellipticals and for 29\% of the galaxies, the classification is unclear.

\subsection{Fibre collisions}

    In the SDSS, for about 6\% of the galaxies the redshifts are not measured due to fibre collisions. The minimum separation between the galaxies (fibres) is 55\arcsec. The distribution of missing galaxies is not uniform in the SDSS, since in overlapping regions, the close neighbours are observed.
    
    To find the galaxies with unmeasured redshifts in our sample, we used the SDSS CAS tables \texttt{sdssTilingInfo} and \texttt{sdssTiledTargetAll}. From these tables, we get the list of all unmeasured galaxies and the tiling group number. The tiling group number and galaxy coordinates are used to find the observed neighbouring galaxies.

    For every unobserved galaxy, we find a closest neighbour in the spectroscopic galaxy sample and raise the missing galaxies flag (\texttt{flagfc}) for that galaxy. Of course, several galaxies in the photometric sample can be close to the same observed galaxy. The flag value \texttt{flagfc} gives the number of missing galaxies close to it.
    
    The missing neighbours can be true neighbours of the galaxy, but they can also be foreground or background galaxies. \citet{Zehavi:02} shows that about 60\% of the galaxies have a redshift close to the neighbour, observed for the redshift. Our visual inspection of such close pairs confirms that about half of the missing neighbours seem to be associated with the observed counterparts.

\begin{figure}
   \centering
   \includegraphics[width=80mm]{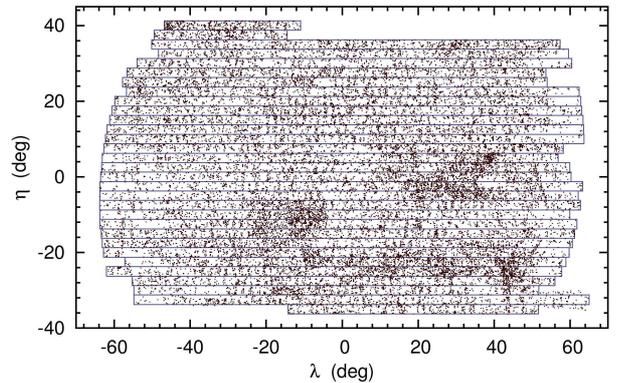}
   \caption{The distribution of the fibre collision galaxies in the SDSS survey. The survey mask used for the density field calculation is shown by grey lines.}
   \label{fig:maskfc}
\end{figure}

    Figure~\ref{fig:maskfc} shows the distribution of fibre collision galaxies in the SDSS sample. The distribution is quite uniform, except in a few regions, where the number of missing galaxies is larger.

%======================================================================
\section{Conclusions and discussion}

    We have constructed a group catalogue for the SDSS DR8 sample, following the same procedure that we used for the DR7 sample \citep{Tago:10}. Since the spectroscopic data is the same in the DR7 and DR8, the new catalogue is similar to the previous DR7 group catalogue. The improvements are in the area of initial galaxy selections and from the SDSS side, the photometry of galaxies.
    
    In addition to the properties, presented in our previous catalogue, we added some new qualitative information. Most importantly, we tracked the missing galaxies in the SDSS (due to fibre collisions) and calculated the environmental density parameters for each galaxy and group. We also added our galaxy morphology as derived in \citet{Tempel:11a}.
    
In the study of galaxy groups, the most important problem at present is the dynamical status of groups of galaxies. Recently \citet{Plionis:06} and \citet{Tovmassian:09} studied shapes and virial properties of groups and found a strong dependence on richness and concluded that groups are not in dynamical equilibrium but rather are at various stages of their virialisation process.

The dynamical status of groups is also characterised by the existence of subgroups  and substructures seen in many studies \citep{Burgett:04, Coziol:09, Einasto:10}. These studies also support the non-virialised nature of groups of galaxies. With both observational and simulated data \citet{Niemi:07} showed that about 20\% of nearby groups are not bound and are groups merely in a visual sense.

    The controversial results  obtained by various authors may be an indication that our present knowledge of groups of galaxies is poor. An optimistic viewpoint is that the study of a broad and inhomogeneous class of galaxy systems -- groups of galaxies -- will help step by step to solve important problems of galaxy formation and evolution, and of the evolution of the large scale structure.
  
\begin{acknowledgements}
      We thank the referee for useful comments that helped to improve the paper. We thank Enn Saar for critically reading the text and for calculating corrected distances for galaxies in groups, and Maret Einasto for useful suggestions. This work was supported by the Estonian Science Foundation grants 7765, 8005, 9428, MJD272 and the Estonian Ministry for Education and Science research projects SF0060067s08. We acknowledge the support by the Centre of Excellence of Dark Matter in (Astro)particle Physics and Cosmology (TK120). E.T. also thanks the University of Valencia (supported by the Generalitat Valenciana project of excellence PROMETEO/2009/064), where part of this study was performed. All the figures have been made using the gnuplot plotting utility.
      
      We are pleased to thank the SDSS-III Team for the publicly available data releases. Funding for the SDSS-III has been provided by the Alfred P. Sloan Foundation, the Participating Institutions, the National Science Foundation, and the U.S. Department of Energy Office of Science. The SDSS-III web site is \texttt{http://www.sdss3.org/}.
      
      SDSS-III is managed by the Astrophysical Research Consortium for the Participating Institutions of the SDSS-III Collaboration including the University of Arizona, the Brazilian Participation Group, Brookhaven National Laboratory, University of Cambridge, University of Florida, the French Participation Group, the German Participation Group, the Instituto de Astrofisica de Canarias, the Michigan State/Notre Dame/JINA Participation Group, Johns Hopkins University, Lawrence Berkeley National Laboratory, Max Planck Institute for Astrophysics, New Mexico State University, New York University, Ohio State University, Pennsylvania State University, University of Portsmouth, Princeton University, the Spanish Participation Group, University of Tokyo, University of Utah, Vanderbilt University, University of Virginia, University of Washington, and Yale University.
\end{acknowledgements}

%\begin{thebibliography}{}
%\end{thebibliography}
%\bibliographystyle{aa}
%\bibliography{group}{}

\appendix
\section{Description of the catalogue}\label{app:cat}

    The catalogue of groups and clusters of galaxies consists of two tables. The first table lists the galaxies that we used to generate our catalogue of groups and clusters, and the second one describes the group properties. Both these catalogues include all the basic entities (distances, coordinates, luminosities, etc) that were present in our previous catalogues, as well the new parameters that are described in this paper.
    
    The catalogues are accessible at \texttt{http://www.aai.ee/$\sim$elmo/dr8groups/} with a complete description in the \texttt{readme.txt} file. We give these catalogues as a \texttt{fits} table with two extensions: one for galaxies and second one for groups. We will also upload the catalogues to the Strasbourg Astronomical Data Center (CDS).

\subsection{Description of the galaxy catalogue}

The galaxy catalogue contains the following information (the column numbers are given in square brackets):
\begin{enumerate}
  \item{[1]\,\texttt{id} --} a unique identification number for galaxies, used by us;
  \item{[2]\,\texttt{idcl} --}  the group/cluster id;
  \item{[3]\,\texttt{nrich} --} the richness of the group the galaxy belongs to;
  \item{[4]\,\texttt{redshift} --} the redshift, corrected to the CMB rest frame;
  \item{[5]\,\texttt{dist} --} the co-moving distance in units of $h^{-1}\,$Mpc (calculated directly from the redshift);
  \item{[6]\,\texttt{distcl} --} the co-moving distance to the group/cluster centre, where the galaxy belongs to, in units of $h^{-1}\,$Mpc, calculated as an average over all galaxies, belonging to the group/cluster;
  \item{[7--8]\,\texttt{ra, dec} --} the right ascension and declination (deg);
  \item{[9--10]\,\texttt{lon, lat} --} the galactic longitude and latitude (deg);
  \item{[11--12]\,\texttt{eta, lam} --} the SDSS survey coordinates $\eta$ and $\lambda$ (deg);
  \item{[13--17]\,\texttt{mag\_$x$} --} the Galactic extinction corrected Petrosian magnitude ($x\in ugriz$ filters);
  \item{[18--22]\,\texttt{absmag\_$x$} --} the absolute magnitude of the galaxy, $k$+$e$-corrected ($x\in ugriz$ filters, in units of $\mathrm{mag}+5\log_{10}h$);
  \item{[23--27]\,\texttt{kcor\_$x$} --} the $k$-correction by the KCORRECT algorithm ($x\in ugriz$ filters);
  \item{[28]\,\texttt{lumr} --} the observed luminosity in the $r$-band in units of $10^{10}h^{-2}L_\odot$, where $M_\odot=4.64$ \citep{Blanton:07};
  \item{[29]\,\texttt{w} --} the weight factor for the  galaxy (\texttt{w}$\cdot$\texttt{lumr} was used to calculate the luminosity density field);
  \item{[30]\,\texttt{rank} --} the galaxy rank in its group, calculated based on the galaxy luminosity: for the most luminous galaxy, the rank is~1;
  \item{[31--35]\,\texttt{den$a$} --} the normalised environmental density of the galaxy for various smoothing scales ($a=1,\,2,\,4,\,8,\,16$~$h^{-1}\,$Mpc); for galaxies outside our survey mask we use the value $-999$, indicating that the galaxy is outside the mask;
  \item{[36]\,\texttt{edgedist} --} the co-moving distance of the galaxy from the border of the survey mask;
  \item{[37]\,\texttt{morf} --} the morphology of the galaxy (0 -- unclear, 1 -- spiral, 2 -- elliptical) as described in Sect.~\ref{sect:morf};
  \item{[38]\,\texttt{hcearly} --} the \citet{Huertas-Company:11} mark (probability) of being an early type galaxy;
  \item{[39]\,\texttt{dr8objid} --} the SDSS DR8 photometric object identification number;
  \item{[40]\,\texttt{dr8specobjid} --} the SDSS DR8 spectroscopic object identification number;
  \item{[41]\,\texttt{iddr7} --} an unique identification number to link with the entries in our previous DR7 group catalogue;
  \item{[42]\,\texttt{zobs} --} the observed redshift (without the CMB correction, as given in the SDSS CAS);
  \item{[43]\,\texttt{distcor} --} the co-moving distance of the galaxy when the finger-of-god effect is suppressed (as used in luminosity density field calculations);
  \item{[44]\,\texttt{flagfc} --} if greater than zero, indicates that the galaxy has a missing 
(with unknown redshift) neighbour due to a fibre collision.
\end{enumerate}

\subsection{Description of the group/cluster catalogue}

The catalogue of groups/clusters contains the following information (the column numbers are given in square brackets):
\begin{enumerate}
    \item{[1]\,\texttt{idcl} -- the group/cluster id;} 
    \item{[2]\,\texttt{nrich} -- the group richness, number of observed galaxies in group;} 
    \item{[3]\,\texttt{zcl} --} the CMB-corrected redshift of group, calculated as an average over all galaxies, belonging to the group/cluster;
    \item{[4]\,\texttt{distcl} --} the co-moving distance to the group centre ($h^{-1}$Mpc);
    \item{[5--6]\,\texttt{racl, deccl} --} the equatorial coordinates, the right ascension and declination (deg);
    \item{[7--8]\,\texttt{loncl, latcl} --} the galactic longitude and latitude (deg);
    \item{[9--10]\,\texttt{etacl, lamcl} --} the SDSS survey coordinates $\eta$ and $\lambda$ (deg);
    \item{[11]\,\texttt{sizesky} --} the maximum linear size of the group in the sky (in physical coordinates, $h^{-1}$Mpc);
    \item{[12]\,\texttt{rvir} --} the virial radius in $h^{-1}$Mpc (the projected harmonic mean, in physical coordinates);
    \item{[13]\,\texttt{sigma\_sky} --} the rms deviation of the projected distance in the sky from the group centre ($\sigma_r$ in physical coordinates, in $h^{-1}$Mpc), $\sigma_r$ defines the extent of the group in the sky;
    \item{[14]\,\texttt{sigma\_v} --} the rms radial velocity deviation ($\sigma_V$ in physical coordinates,  in \mbox{km\,s$^{-1}$});
    \item{[15]\,\texttt{lumobs\_r} --} the observed luminosity, the sum of observed galaxy luminosities ($10^{10}h^{-2}L_\odot$);
    \item{[16]\,\texttt{lumtot\_r} --} the estimated total luminosity of the group ($10^{10}h^{-2}L_\odot$), the observed luminosity multiplied by a weight factor;
    \item{[17]\,\texttt{dlink} --} the transverse co-moving linking length used to define the group ($h^{-1}$Mpc). The ratio between the radial and the transversal linking lengths is taken to be 10 (the radial linking length in km\,s$^{-1}$ is transformed into a formal ``distance'' in $h^{-1}$Mpc);
    \item{[18--22]\,\texttt{den$a$}cl --} the environmental density for the group for various smoothing scales ($a=1,\,2,\,4,\,8,\,16$~$h^{-1}$Mpc), averaged over group galaxies inside the survey mask;
    \item{[23]\,\texttt{edgedistcl} --} the minimum co-moving distance of group galaxies from the edge of the survey mask. Zero, if at least one member is outside the survey mask.
    \item{[24]\,\texttt{flagfccl} --} if greater than zero, indicates that the group contains potentially missing (with unknown redshift) galaxies due to the fibre collisions.
\end{enumerate}

\end{document}